\documentclass[twoside,fleqn]{article}
\usepackage{espcrc2}
\usepackage{graphicx}
\newcommand{\be}{\begin{equation}}
\newcommand{\ee}{\end{equation}}
\newcommand{\chiPT}{$\chi$PT}

\title{ 
\vspace{-2.6cm}
\hfill \rm \null \hfill
\hbox{\normalsize ADP-03-128/T564} \\
\vspace{-2mm}
\hfill \hbox{\normalsize DESY 03-157} \\
\vspace{1.65cm}
Electromagnetic Form Factors with FLIC fermions}

\author{J.~M.~Zanotti\address[CSSM]{Special Research Center for the
    Subatomic Structure of Matter, and		\\
    Department of Physics, University of Adelaide
    Adelaide SA 5005  Australia}\address[DESY]{John von
    Neumann-Institut f\"ur Computing
    NIC, \\
    Deutsches Elektronen-Synchrotron DESY, D-15738 Zeuthen, Germany}
    \thanks{Presented by J.~M.~Zanotti at Lattice '03},
  D.~B.~Leinweber\addressmark[CSSM], A.~G.~Williams\addressmark[CSSM]
  and J.~B.~Zhang\addressmark[CSSM] }

\begin{document}

\begin{abstract}
  The Fat-Link Irrelevant Clover (FLIC) fermion action provides a new
  form of nonperturbative ${\cal O}(a)$ improvement and allows
  efficient access to the light quark-mass regime.  FLIC fermions
  enable the construction of the nonperturbatively ${\cal
  O}(a)$-improved conserved vector current without the difficulties
  associated with the fine tuning of the improvement coefficients.
  The simulations are performed with an ${\cal O}(a^2)$ mean-field
  improved plaquette-plus-rectangle gluon action on a $20^3\times 40$
  lattice with a lattice spacing of 0.128 fm, enabling the first
  simulation of baryon form factors at light quark masses on a large
  volume lattice.
  Magnetic moments, electric charge radii and magnetic radii are
  extracted from these form factors, and show interesting chiral
  nonanalytic behavior in the light quark mass regime.
\end{abstract}

\maketitle



\section{INTRODUCTION}

The magnetic moments of baryons have been identified
\cite{Leinweber:2001jc,Leinweber:2002qb} as providing an excellent
opportunity for the direct observation of chiral nonanalytic behavior
in lattice QCD, even in the quenched approximation.  This paper will
present results for baryon electromagnetic structure in which the
chiral nonanalytic behaviour predicted by quenched chiral perturbation
theory is observed in the numerical simulation results.


%

FLIC fermions provide a new form of nonperturbative ${\cal O}(a)$
improvement \cite{Leinweber:2002bw,inPrep} where near-continuum
results are obtained at finite lattice spacing.  Access to the light
quark mass regime is enabled by the improved chiral properties of the
lattice fermion action \cite{inPrep}.
%
%
A key feature of the FLIC fermion approach is that the fine tuning of
${\cal O}(a)$-improvement term coefficients is not necessary.  The
improvement terms are irrelevant operators which are constructed using
fat-links.  Remaining perturbative renormalizations are accurately
accounted for by small mean-field improvement corrections.  Hence, we
are able to determine the form factors of octet and decuplet baryons
with unprecedented accuracy.

\vspace*{-0.1cm}
\section{LATTICE ACTIONS}
\label{FLinks}

The simulations are performed using a mean-field ${\cal
O}(a^2)$-improved Luscher-Weisz \cite{Luscher:1984xn} gauge action on
a $20^3 \times 40$ lattice with a lattice spacing of 0.128 fm as
determined by the Sommer scale $r_0=0.50$ fm.  We use a minimum of 255
configurations and the error analysis is performed by a third-order,
single-elimination jackknife.

Fat links 
are created using APE smearing \cite{ape} followed by projection of
the smeared link back to $SU(3)$.  We select a smearing fraction of
$\alpha = 0.7$ (keeping 0.3 of the original link) and iterate the
process six times \cite{Bonnet:2000dc}.  Further details of FLIC
fermion actions can be found in Ref.~\cite{FATJAMES}.

For fat links, the mean link $u_0 \approx 1$, enabling the use of
highly improved definitions of the lattice field strength tensor,
$F_{\mu\nu}$ \cite{Bilson-Thompson:2002jk}.  In particular, we employ
an ${\cal O}(a^4)$-improved definition of $F_{\mu\nu}$ in which the
standard clover-sum of four $1 \times 1$ Wilson loops lying in the
$\mu ,\nu$ plane is combined with $2 \times 2$ and $3 \times 3$ Wilson
loop clovers.  Moreover, mean-field improvement of the coefficients of
the clover and Wilson terms of the fermion action is sufficient to
accurately match these terms and eliminate ${\cal O}(a)$ errors from
the fermion action.  As the conserved vector current has its origin in
the fermion action, mean-field improvement of the irrelevant operator
terms is also sufficient to accurately remove ${\cal O}(a)$ errors,
providing a nonperturbatively ${\cal O}(a)$-improved conserved vector
current.

For the construction of the ${\cal O}(a)$-improved conserved vector
current, we follow the technique proposed by Martinelli {\it et al.}
\cite{Martinelli:ny}.  The standard conserved vector current for
Wilson-type fermions is derived via the Noether procedure
\begin{eqnarray}
j_\mu^{\rm C} &\equiv& \frac{1}{4}\bigl[\overline{\psi}(x) (\gamma_\mu -
r)U_\mu(x) \psi(x+\hat{\mu}) \nonumber \\ 
&+& \overline{\psi}(x+\hat{\mu}) (\gamma_\mu + r)U_\mu^\dagger(x)
\psi(x) \nonumber \\
&+& (x\rightarrow x-\hat{\mu})\bigr] .
\label{conserved}
\end{eqnarray}
The ${\cal O}(a)$ improvement term is also derived from the fermion
action and is constructed in the form of a total four-divergence,
preserving charge conservation.  The ${\cal O}(a)$-improved conserved
vector current is
\be
j_\mu^{\rm CI} \equiv j_\mu^{\rm C} (x) + \frac{r}{2} C_{CVC}\, a \sum_\rho
\partial_\rho \bigl( \overline{\psi}(x) \sigma_{\rho\mu}\psi(x)\bigr)
\, ,
\label{impconserved}
\ee
where $C_{CVC}$ is the improvement coefficient for the conserved
vector current and we define
\be
\partial_\rho \bigl( \overline{\psi}(x) \psi(x)\bigr) \equiv
\overline{\psi}(x) \bigl( \overleftarrow{\nabla}_\rho +
\overrightarrow{\nabla}_\rho \bigr) \psi(x)\, .
\ee
The terms proportional to the Wilson parameter $r$ in
Eq.~(\ref{conserved}) and the four-divergence in
Eq.~(\ref{impconserved}) have their origin in the irrelevant operators
of the fermion action and vanish in the continuum limit.
Nonperturbative improvement is achieved by constructing these terms
with fat-links.  As we have stated, perturbative corrections are small
for fat-links and the use of the tree-level value for $C_{CVC} = 1$
together with small mean-field improvement corrections ensures that
${\cal O}(a)$ artifacts are accurately removed from the vector
current.  This is only possible when the current is constructed with
fat-links.  Otherwise, $C_{CVC}$ needs to be appropriately tuned to
ensure all ${\cal O}(a)$ artifacts are removed.


A fixed boundary condition at $t=0$ is used for the fermions and
gauge-invariant Gaussian smearing \cite{Gusken:qx,Zanotti:2003fx} in
the spatial dimensions is applied at the source at $t=8$ to increase
the overlap of the interpolating operators with the ground state while
suppressing excited state contributions.  The technique used for
constructing the three-point functions follows the procedure outlined
in detail in Refs.~\cite{Leinweber:1990dv,Leinweber:1992hy}. In
particular, we use the sequential source technique at the current
insertion.  Correlation functions are made purely real with exact
parity through the consideration of $U$ and $U^*$ link configurations.
Electric and magnetic form factors are extracted by constructing
ratios of two- and three-point functions.
We simulate at the smallest finite $q^2$ available on our lattice,
$ {\vec q = \frac{2 \pi}{a L} \widehat x}$.

\section{RESULTS}
\label{discussion}

\begin{figure}[t]
\begin{center}
{\includegraphics[height=\hsize,angle=270]{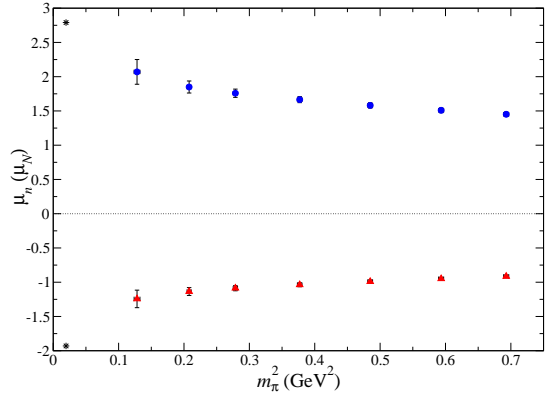}}
\vspace*{-1.0cm}
\caption{FLIC fermion simulation results for the magnetic moments of
  the proton ($\circ$) and neutron ($\triangle$) in
  quenched QCD. }
\label{PNmoments}
\end{center}
\vspace{-1.0cm}
\end{figure}

Figure \ref{PNmoments} displays FLIC fermion simulation results for the
magnetic moments of the proton and neutron in quenched QCD.
At heavy quark masses we note that the magnetic moments of both
nucleons display linear behaviour when plotted as a function of
$m_\pi^2$.  As one approaches the light quark mass regime, we find
evidence of non-analytic behaviour in the nucleon magnetic moments as
predicted by quenched \chiPT\ \cite{Leinweber:2001jc,Leinweber:2002qb}. 
In fact, if we were to flip the sign of the neutron magnetic moment,
we would find that the proton and neutron have a very similar
behaviour as a function of $m_\pi^2$ in the light quark mass regime as
predicted by the leading non-analytic contributions of quenched \chiPT.

\begin{figure}[t]
\begin{center}
{\includegraphics[height=\hsize,angle=270]{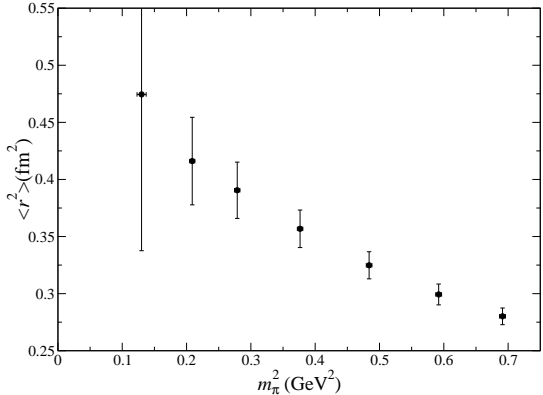}}
\vspace*{-1.0cm}
\caption{FLIC fermion simulation results for the charge radius of the
  proton in quenched QCD. }
\label{chargeRadii}
\end{center}
\vspace{-1.0cm}
\end{figure}

Figure \ref{chargeRadii} displays results for the proton charge radius
obtained from a dipole form factor ansatz.  Some curvature is
emerging as the chiral limit is approached.

\begin{figure}[t]
\begin{center}
{\includegraphics[height=\hsize,angle=270]{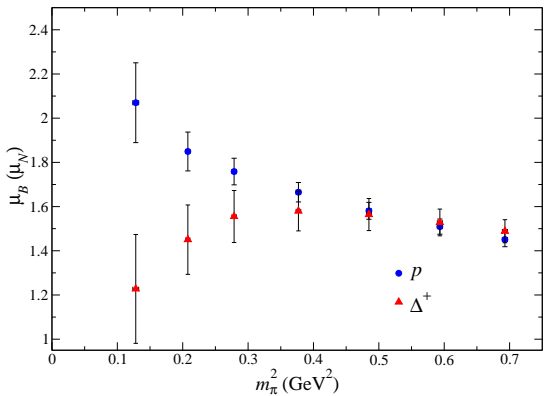}}
\vspace*{-1.0cm}
\caption{FLIC fermion simulation results for the magnetic moments of
  the proton ($\circ$) and $\Delta^+$ resonance ($\triangle$) in
  quenched QCD. }
\label{PDmoments}
\end{center}
\vspace{-1.0cm}
\end{figure}

Figure \ref{PDmoments} displays FLIC fermion simulation results for the
magnetic moments of the proton and $\Delta^+$ resonance in quenched QCD.
At large pion masses, the $\Delta$ moment is enhanced relative to the
proton moment in accord with earlier lattice QCD results
\cite{Leinweber:1990dv,Leinweber:1992hy} and model expectations.
However as the chiral regime is approached the nonanalytic behavior of
the quenched meson cloud is revealed, enhancing the proton and
suppressing the $\Delta^{+}$ in accord with the expectations of
quenched \chiPT\ \cite{Leinweber:2003ux,RossLat}.  The quenched
artifacts of the $\Delta$ provide an unmistakable signal for the onset
of quenched chiral nonanalytic behavior.

We have presented the first lattice QCD simulation results for the
electromagnetic form factors of the nucleon and $\Delta$ at quark
masses light enough to reveal unmistakable quenched chiral
nonanalytic behavior.

\vspace*{0.3cm} This work was supported by the Australian Research
Council.  We thank the Australian Partnership for Advanced
Computing (APAC) for generous grants of supercomputer time which have
enabled this project.


\vfill
\end{document}